\documentclass[a4paper,11pt]{article}

\usepackage{jheppub} 

\usepackage[T1]{fontenc} 

\usepackage{caption,subcaption}
\usepackage{tikz}
\usetikzlibrary{calc}
\usetikzlibrary{arrows,decorations.pathmorphing,patterns}
\usepackage{soul}

\numberwithin{equation}{section}
\usepackage{version}
\usepackage{mathrsfs} 

\usepackage{hyperref}
\hypersetup{
	colorlinks=true,
	linkcolor=blue,
	filecolor=magenta,      
	urlcolor=blue,
	citecolor=magenta
}

\def\pr{\partial}

\def\be{\begin{equation}}
\def\ee{\end{equation}}
\def\bea{\begin{eqnarray}}
\def\eea{\end{eqnarray}}
\def\ba{\begin{array}}
\def\ea{\end{array}}
\def\bec{\begin{center}}
\def\ec{\end{center}}

\def\a{\alpha} 
\def\b{\beta}  
\def\g{\gamma} 
\def\G{\Gamma}
\def\d{\delta} 
\def\D{\Delta}
\def\e{\epsilon} 
\def\ve{\varepsilon}

\def\th{\theta}

\def\l{\lambda}
\def\L{\Lambda}
\def\m{\mu}
\def\n{\nu}

\def\r{\rho}
\def\s{\sigma}

\def\t{\tau}
\def\f{\phi}
\def\F{\Phi}
\def\vf{\varphi}

\def\o{\omega}

\def\cD{{\cal D}}

\def\cO{{\cal O}}

\def\ta{{\tilde{A}}}
\def\tal{{\tilde{\alpha}}}

\def\bz{\bar{z}}
\def\vk{{\vec k}}

\def\vz{{\vec z}}
\def\vw{{\vec w}}

\title{Double-Copy Supertranslations}%
\author[a]{Pietro Ferrero,}
\author[b]{Dario Francia,}
\author[c]{Carlo Heissenberg}
\author[b]{and Matteo Romoli}

\affiliation[a]{Simons Center for Geometry and Physics, SUNY, Stony Brook, NY 11794, USA}
\affiliation[b]{Roma Tre University and INFN Sezione di Roma Tre, via della Vasca Navale, 84 I-00146 Roma, Italy}
\affiliation[c]{School of Mathematical Sciences, Queen Mary University of London, Mile End Road, London, E1 4NS, UK}

\emailAdd{pferrero@scgp.stonybrook.edu}
\emailAdd{dario.francia@uniroma3.it}
\emailAdd{c.heissenberg@qmul.ac.uk}
\emailAdd{matteo.romoli@uniroma3.it}

\abstract{In the framework of the convolutional double copy, we investigate the asymptotic symmetries of the gravitational multiplet stemming from the residual symmetries of its single-copy constituents at null infinity. We show that the asymptotic symmetries of Maxwell fields in $D=4$ imply ``double-copy supertranslations’’, {\it i.e.} BMS supertranslations and two-form asymptotic symmetries, together with the existence of infinitely many conserved charges involving the double-copy scalar. With the vector fields in  Lorenz gauge, the double-copy parameters display a radial expansion involving logarithmic subleading terms, essential for the corresponding  charges to be nonvanishing.}

\notoc

\begin{document}

\maketitle

\vspace{10cm}

\section{Introduction}

The double copy (DC) is a powerful formal tool that, in its most basic incarnation, captures key properties of gravity by suitably combining two copies of a gauge theory. This idea is at the basis of outstanding achievements in scattering amplitude calculations \cite{Bern:2008qj,Bern:2010ue} and also applies, properly reinterpreted, to classical solutions, either exact or perturbative \cite{Monteiro:2014cda, Luna:2015paa, Goldberger:2016iau,Luna:2016hge}. (See \cite{Bern:2019prr, Borsten:2020bgv, Bern:2022wqg, Adamo:2022dcm} for recent reviews and extended references.) It is therefore natural to wonder whether it also links the asymptotic symmetries of gravity and gauge theories, which characterise their infrared structure and lie at the heart of soft theorems and memory effects \cite{Strominger:2017zoo}.

In this work we give a positive answer to this question by showing that, starting from spin-one large gauge transformations, the DC allows one to derive Bondi--Metzner--Sachs (BMS) supertranslations at null infinity, together with  the asymptotic symmetries of two-form gauge fields. Furthermore, one also finds  an infinite set of conserved charges for the scalar field, reproducing those originally found in \cite{Campiglia:2017dpg}, whose rationale in this setting is a consequence of the scalar field being part of a multiplet together with graviton and two-form. A DC perspective on asymptotic symmetries can be found in \cite{Campiglia:2021srh}, with reference to  the self-dual sectors of gauge and gravitational theories. Further related works are \cite{Adamo:2021dfg, Godazgar:2021iae}.

The convolutional approach to the DC was developed in \cite{Anastasiou:2014qba,LopesCardoso:2018xes,Anastasiou:2018rdx, Luna:2020adi, Borsten:2020xbt} and further explored in \cite{Borsten:2019prq, Borsten:2020zgj, Ferrero:2020vww, Beneke:2021ilf, Borsten:2021zir, Godazgar:2022gfw, Liang:2023zxo} from an off-shell perspective. Its original goal was to investigate the meaning of the DC itself at the level of symmetries, and in  this sense it provides a natural framework for our purposes.  In \cite{Ferrero:2020vww} in particular it was implemented at the Lagrangian level up to the cubic order in the Maxwell-like setup of \cite{Francia:2010qp,Campoleoni:2012th, Francia:2016weg}, which  delivers a simple off-shell encoding of the degrees of freedom of the full gravitational multiplet, including the scalar particle. This is achieved by recognising that a gauge-invariant scalar can be encoded off-shell in the trace of the graviton upon restricting the gauge symmetries of the latter to linearised volume-preserving diffeomorphisms. The corresponding action provides a generalisation of unimodular gravity allowing one to encompass in a single tensor the degrees of freedom of the graviton, the scalar and the two-form field. The corresponding Lagrangian description bears a direct connection with the rank-two sector of free tensionless strings \cite{Bengtsson:1986ys, HT,Sagnotti:2003qa,Francia:2004lbf}.

We thus start from the linear Lagrangian equations of \cite{Ferrero:2020vww} in $D=4$ and consider both vectors in Lorenz gauge, so that all single-copy fields, including the spectator scalar entering the definition \eqref{H} of the DC field, satisfy the massless Klein--Gordon equation. In this gauge, in order to identify  non-vanishing asymptotic charges one has to allow for a polyhomogeneous expansion of the spin-one gauge parameters in the radial coordinate, involving logarithmic terms  at first subleading order \cite{Campiglia:2016hvg,Campoleoni:2019ptc}.

 Given these inputs, we apply the method of regions \cite{Beneke:1997zp,Smirnov:2002pj} in order to evaluate the falloffs of the  DC multiplet and of the corresponding parameters at null infinity. As a first output, we find in particular for both the gravitational parameter and the two-form parameter polyhomogeneous radial scalings. For the  graviton,  we recover in this way BMS supertranslations in de Donder gauge as discussed in \cite{Avery:2015gxa,Campiglia:2016efb,Himwich:2019qmj}, with the corresponding  asymptotic  behaviours. Differently, for the two-form in Lorenz gauge, to our knowledge logarithmic falloffs for the parameters were not considered before, and in this sense the output of the DC in this sector provides an original description of the corresponding asymptotic symmetries. The main upshot of our analysis is that such logarithmic falloffs of the parameters  turn out to be actually necessary in order for the corresponding two-form asymtptotic charge to be different from zero. The reason is the same one that requires the presence of similar falloffs  in the spin-one and spin-two cases: namely, if one assumes pure power-law radial behaviour of the gauge parameters, in Lorenz gauge for vector fields or in de Donder gauge for gravitons, the relevant coefficients entering the  asymptotic charges actually vanish, due to the d'Alembert equation that the parameters themselves have to obey in the corresponding gauges. The same mechanism applies to the two-form in Lorenz gauge, although it was seemingly overlooked so far. In this sense, it is notable that the DC automatically encodes the generality of the mechanism and prescribes the proper logarithmic behaviour for the parameters of the multiplet. 
 
 Furthermore, as anticipated, our DC derivation also provides a novel interpretation for the asymptotic charges involving massless scalar fields. While their symmetry origin may be traced back to the massless scalar in $D=4$ being dual to a two-form gauge field \cite{Campiglia:2018see, Francia:2018jtb}, in the DC setup they emerge as a consequence of the existence of supertranslation charges involving a traceful graviton, whose trace actually defines a gauge-invariant and physical degree of freedom.

The paper is organised as follows: in section \ref{cdc} we introduce the basics of the convolutional DC and in section \ref{dc_field_null} we analyse the behaviour of the DC field at null infinity as an output of the corresponding behaviour of its single-copy constituents, while in section \ref{dc_parameters_null} we repeat the analysis for the gauge parameters. In section \ref{dc_asymptotica} we study the asymptotic symmetries of the DC field and interpret them in terms of the individual particles of the corresponding multiplet; in particular, in the spin-two sector we recover in section \ref{dc_parameters_null_symm} both the supertranslation charges and the asymptotic charges involving the scalar field, here identified in the sector involving the trace of the DC field. In section \ref{dc_parameters_null_asymm} we analyse the two-form sector and observe the relevance of the polyhomogeneous expansion of the parameters. Section \ref{conclusions} collects our conclusions and a summary of future directions.

\section{Convolutional double copy} \label{cdc}

The  DC field 
\begin{align}  \label{H}
H_{\, \m \n} \, &= \,  A^{a}_{\m} \circ \Phi^{-1}_{aa'} \circ \tilde{A}^{a'}_{\n} \, = \, A_{\m}\,\star\, \ta_{\n}
\end{align} 
is defined as a double convolution in Cartesian coordinates, where 
%
\be  \label{convolution}
[f\circ g](x) = \int d^Dy f(y)g(x-y)
\ee
denotes the standard convolution,
$A^{a}_{\m}$ and $\tilde{A}^{a'}_{\n}$ are gauge fields taking values in two generically different Lie algebras, so that $a$ and $a'$ represent color indices in the adjoint representation, while $\Phi^{-1}_{aa'}$ is the convolution inverse of a scalar field in the biadjoint of the two algebras\footnote{In the spirit of \cite{Anastasiou:2014qba}, the global transformations of $A^{a}_{\m}$ and $\tilde{A}^{a'}_{\n}$ are compensated by those of $\Phi^{-1}_{aa'}$, while we only consider the linearised limit of the respective local transformations.} defined by 
\begin{equation}\label{conv_inv}
\F_{aa'} \circ \F^{-1}_{bb'} (x) = \d_{ab} \d_{a'b'} \d^{(D)} (x)\, .
\end{equation}
The DC field encodes the degrees 
of freedom of a graviton, a Kalb--Ramond particle and a massless scalar, whose off-shell fields in $D=4$ can be identified in terms of the following projections \cite{LopesCardoso:2018xes, Ferrero:2020vww}
\begin{equation} \label{DC multiplet}
\begin{split}
h_{\mu\nu} \,&=\, H^S_{\mu\nu}  -  \frac{1}{2}\, \eta_{\m\n} \, \vf, \\
B_{\mu\nu} &=\, H^A_{\mu\nu}, \\
\vf  &=\, H^{\a}_{\a}  -  \frac{\pr^{\, \a} \pr^{\, \b} H_{\, \a \b}}{\Box},
\end{split}
\end{equation}
where $H^S_{\mu\nu}$ and $H^A_{\mu\nu}$ denote  symmetric and antisymmetric parts of $H_{\m\n}$, respectively.  The Abelian gauge transformation of the DC field, encoding the local symmetries of the graviton and of the two-form field, 
\begin{equation} \label{gaugeH}
\d H_{\m\n} \, = \, \pr_{\m} \a_\n + \pr_{\n} \tilde{\a}_{\m},
\end{equation}
involves the parameters
\begin{equation} \label{DC parameters}
		\a_{\m} = \e \star \tilde{A}_\m, \qquad  \tilde{\a}_{\m} = A_\m \star \tilde{\e},
\end{equation}
where $\e^a$ and $\tilde{\e}^{a'}$ are  Lorentz scalars parameterising the gauge transformations of the spin-one gauge fields:
\begin{equation}
\d A_\m^a=\partial_\m\e^a\,, \qquad 
\d\tilde{A}_\m^{a'}=\partial_\m\tilde{\e}^{a'}\,.
\end{equation}
In particular, in the convolutional DC dictionary the reducibility of the gauge symmetry of the two-form $H^A_{\m\n}$, {\it i.e.} its gauge-for-gauge transformations, corresponds to DC parameters involving pure-gauge vectors: 
\begin{equation} \label{gfg}
A_{\m}=\pr_{\m} \r, \, \tilde{A}_{\m}=\pr_{\m} \tilde{\r} \;\;\rightarrow \;\; \a_{\m} - \tilde{\a}_\m =  \pr_{\m} \left(\e * \tilde{\r} - \r * \tilde{\e}\right). 
\end{equation}
The Lagrangian 
\begin{equation} \label{lagrangian}
    {\cal L} \, = \, \frac{1}{2} \, H^{\mu\nu} \big(\eta_{\mu \rho} \eta_{\nu \sigma} \Box - \eta_{\mu \rho}  \partial_{\nu} \partial_{\sigma} - \eta_{\nu \sigma} \partial_{\mu} \partial_{\rho}  \big) \, H^{\rho \sigma} 
\end{equation}
is gauge invariant under \eqref{gaugeH} provided the parameters satisfy
\begin{equation} \label{tdiff}
\partial^{\mu} (\alpha_{\mu} + \tilde{\alpha}_{\mu}) = 0\, ,
\end{equation}
and encodes the free propagation of the full DC multiplet. Equivalent Lagrangians whose gauge invariance holds without the need for the transversality condition \eqref{tdiff} involve either additional fields or the curvature for the DC field. All these options were presented in \cite{Ferrero:2020vww}. 

In the ensuing sections, we will study the asymptotic behaviour at null infinity of the fields defined in \eqref{DC multiplet} and of the gauge parameters in \eqref{DC parameters}, assuming that their single-copy constituents  satisfy the asymptotic conditions granting the existence of  spin-one asymptotic symmetries in Lorenz gauge \cite{Campiglia:2016hvg,Campoleoni:2019ptc}. 

\section{Double-copy field at null infinity} \label{dc_field_null}

In this section we consider the expansion of the fields involved in the DC construction at null infinity. In particular, we shall assume a given fall-off behaviour for the single-copy fields and work out the implications of this behaviour for the DC field $H_{\m\n}$. The main technical step is to understand the interplay between the expansion of the fields for large values of the radial coordinate $r$ and the operation of convolution.

Choosing the Lorenz gauge for the spin-one fields\footnote{In the setup defined by \eqref{lagrangian} with DC parameters related by \eqref{tdiff} one can enforce the Lorenz gauge on one vector and deduce the same condition on the second vector as a consequence of \eqref{tdiff}. Alternatively, starting from the equivalent unconstrained setup also presented in \cite{Ferrero:2020vww}, one can directly enforce the Lorenz gauge on both vectors and reduce the equations of motion to \eqref{eom}.}, $\partial^{\m } A_{\m}  = 0 = \partial^{\m } \tilde{A}_{\m}$, the DC field satisfies 
\begin{equation} \label{LorenzH}
 \partial^{\m} H_{\m \n} = 0 =  \partial^{\n} H_{\m \n}\, ,  
\end{equation}
so that its free Lagrangian equations reduce to \cite{Ferrero:2020vww}
\begin{equation} \label{eom}
\Box H_{\m\n} = 0\, .
\end{equation}
We assume all single-copy fields to be  on shell, and write the solutions to their equations of motion as follows\footnote{Let us note that, even in the presence of localised sources, the homogeneous solutions \eqref{A,Phi exp} provide good approximations for the fields far from the sources \cite{Cristofoli:2021vyo}.}
\begin{equation}\label{A,Phi exp}
		\begin{aligned}
			A_{\m a}(x) =& \int_k e^{ik \cdot x} a^{(A)}_{a i}(k) \,\varepsilon^i_\m(k)\,,\\
			\tilde{A}_{\n a'}(x) =& \int_k e^{ik\cdot x} \tilde{a}^{(A)}_{a' j}(k) \,\tilde{\varepsilon}^j_\n(k)\,,\\
			\F_{aa'}(x) =& \int_k e^{ik\cdot x} a^{(\F)}_{aa'}(k)\,, \\
		\end{aligned}
\end{equation}
where
\begin{equation}
\int_kf(k)=\int \frac{d^4k}{(2\pi)^4} \theta(k^0) \d(k^2) f(k)+\text{c.c.}\, .
\end{equation}
In addition, we also assume that the leading-order radial falloffs of all single-copy fields be radiation falloffs (tantamount to Coulombic 
falloffs in $D=4$), and thus such that the Cartesian components in \eqref{A,Phi exp} all scale as follows,
\begin{equation}
    A_{\mu} \sim \mathcal{O}\left(\frac{1}{r}\right) ,
\end{equation}
and similarly for $\tilde{A}_{\m}$ and $\Phi$.

The general solution to \eqref{eom} is 
\begin{equation}\label{H_Fourier}
	H_{\m\n} = \int_k e^{ik\cdot x} a^{(H)}_{ij} (k) \ve_{\m\n}^{ij}(k)\,.
\end{equation}
On the other hand, taking the Fourier transform of Eq.~\eqref{H}, which relates convolutions to ordinary products, by the definition of the convolutional inverse\footnote{
We employ \eqref{conv_inv} 
to identify
\begin{equation*}
\int_p a^{(A)}_i (p) \F^{-1}_{bb'} (p) [\ \cdot \ ]  =  \int \frac{d^4p}{(2 \pi)^4}\frac{a^{(A)}_i (p)}{a^{(\F)}(p)} [\ \cdot \ ] \, ,
\end{equation*}
where 
$\F^{-1}_{bb'} (p)$ denotes the Fourier coefficients of $\F^{-1}_{bb'} (x)$.} 
 we eventually find
\begin{equation} \label{H parameters}
a^{(H)}_{ij} = \frac{a^{(A)}_i \tilde{a}^{(A)}_j}{ a^{(\F)}}\,,  \qquad \ve_{\m\n}^{ij} = {\ve^i_\m\tilde{\ve}^j_\n }\, ,
\end{equation}
where in the previous formula and in the following ones color indices are implicit. We then determine the falloffs of $ H_{\m\n}$ according to the method of regions \cite{Beneke:1997zp, Smirnov:2002pj}.
To this end,  we parameterise  Minkowski coordinates and momenta in terms of Bondi coordinates as (see appendix \ref{coordinates}  for details)
\begin{equation} \label{Bondi parameter}
\begin{split}
	&x^\m = u t^\m + \frac{2r}{1+|\vz\,|^2}\,q^\m(\vz\,)\, ,\\&
	k^\m  = \mu t^\m + \o q^\m(\vw)\, .
\end{split}
\end{equation}

Changing  integration variables in \eqref{H_Fourier}, we get 
\begin{equation} \label{Hfourier}
\begin{split}
	H_{\m\n} =&\int_0^{\infty} d \omega \frac{\omega}{2} \int d^2 {\vw} \left[e^{-i \omega q^0(\vw) u-\frac{i \omega r|{{\vw}}-{\vz}|^2}{1+|\vz\,|^2}} a^{(H)}(\omega q(\vw)) \, \ve_{\m\n}(q(\vw)) + \text{c.c.}\, \right] ,
\end{split}
\end{equation}
where in particular the Minkowskian components of the polarisation tensor are  treated as functions of  $\vw$.

In the large-$r$ limit, the leading behaviour of $H_{\m\n}$ is determined by those regions where
\begin{equation}
	\frac{\omega r |{{\vw}}-{\vz}\,|^2}{1+|\vz\,|^2} \sim {\cal O} (1)\, . 
\end{equation}
There are two relevant regions: $|\vw-\vz\,|^2 \sim \tfrac{1}{r}$ (collinear region) and $\o \sim \tfrac{1}{r}$ (regular region),  whereas contributions from other regions are either scale-less  or lead to rapidly oscillating integrals.

The collinear region is the one responsible for the radiative order $H_{\m \n} \sim \mathcal{O}(\tfrac{1}{r})$. Indeed, by performing the change of variables  $\vw = \vz + \tfrac{\vec{s}}{\sqrt{r}}$ with $\vec{s}$ formally of $\cO(1)$ and expanding around $\vz$, one obtains a Gaussian integral in $\vec{s}$ to leading order so that \eqref{Hfourier} gives 
\begin{equation}\label{H_coll}
    H^\text{coll}_{\m\n} \sim \frac{i\pi}{r}\int_0^\infty \frac{d\o}{2\o} e^{-i\o q^0(\vz\,)u}a^{(H)}(\o q(\vz\,)) \ve_{\m\n}(q(\vz\,)) + \text{c.c.} \, .
\end{equation}
The radiative order is present for all fields
satisfying the massless wave equation (at least asymptotically) and in this respect its DC origin is to be traced back to the fact that whenever the vectors are on-shell and in the Lorenz gauge the DC field satisfies indeed  $\Box H_{\m\n} = 0$ \cite{Ferrero:2020vww}. 

In the regular region, on the other hand, one has to take into account the expansion of the Fourier coefficients for small frequency. 
Upon assuming the following leading-order behaviour   (denoted with the subscript $0$) 
\begin{equation}\label{eq:aHsimaH0}
   a^{(H)}(\omega q(\vw)) \sim \o^{\b-2} a^{(H)}_0(q(\vw)), 
\end{equation}
where $\b > 0$ for the integral in \eqref{Hfourier} to be convergent,  in general one would end up with 
\begin{equation}\label{H_reg}
    H^\text{reg}_{\m\n} \sim \frac{\G(\b)}{(ir)^\b}(1+|\vz\,|^2)\int d^2\vw\,\frac{a^{(H)}_0(q(\vw)) \ve_{\m\n}(q(\vw))}{2|\vw-\vz\,|^{2\b}}+ \text{c.c.}\,,
\end{equation}
where $\G(\b)$ is the gamma function. We thus obtain
contributions ${\cal{O}}( \tfrac{1}{r^{\b}})$ from this region, which would be overleading with respect to radiation for $\b < 1$.  However, if the single-copy constituent fields in \eqref{H} all  fall off like $\frac{1}{r}$ at null infinity, by means of a similar reasoning one finds
\begin{equation}\label{single_copy_scaling}
    a^{(A)} \sim \frac{1}{\o}a^{(A)}_0,\qquad\; a^{(\tilde A)} \sim \frac{1}{\o}\tilde{a}^{(\tilde A)}_0, \qquad\;  a^{(\F)} \sim \frac{1}{\o}a^{(\F)}_0,
\end{equation}
or, equivalently,
\begin{equation} \b_A=\b_{\tilde A}=\b_\F = 1 \ .
\end{equation} 
From the first of \eqref{H parameters} we derive 
\begin{equation}
    \b_H = \b_A+\b_{\tilde{A}}-\b_{\F}=1,
\end{equation}
and therefore
\begin{equation}
    a^{(H)}(\omega q(\vw)) \sim \frac{1}{\o}\, a_0^{(H)}(q(\vw))\,,
\end{equation}
consistently with the form of the leading soft factors as determined by the soft theorems \cite{Strominger:2014pwa,Saha:2019tub,Sahoo:2020ryf,Sahoo:2021ctw}, as we shall discuss in the section \ref{dc_asymptotica}. This shows that, for single-copy fields on their mass-shell with radiation leading falloffs, and for vectors in Lorenz gauge, the DC field \eqref{H} is in the same gauge and displays the same leading falloffs  $\sim {\cal O}(\tfrac{1}{r})$. Note that the polarisation vectors and tensors in \eqref{H parameters} are invariant under rescalings of the massless momentum.

To conclude this section we would like to comment on the structure of the leading components of the DC field (similar considerations apply to the vectors as well). To begin with, the regular region cannot provide any $u$-dependence to leading order,  as one can read from \eqref{H_reg}. Furthermore, using
\begin{equation} \label{beta_to_one}
    \cos\left({\frac{\pi}{2} \beta}\right) |\vw-\vz\,|^{-2\beta} = \frac{\pi^{2}}{2} \d^{(2)}(\vw-\vz\,) + \cO(\beta - 1)
\end{equation}
(see {\it e.g.} \cite{Donnay:2022ijr}) one can perform the integral \eqref{H_reg}\footnote{In order to evaluate \eqref{H_reg} via \eqref{beta_to_one} we also used the reality of $a_0^{(H)}$, inherited from the corresponding condition on its single-copy constituents.} in the limit $\b\to 1$ and see that the leading contribution from the regular region gets localised in $\vec{z}$. 

For the collinear region, we see that it captures the contribution of radiation, since, as we see from \eqref{H_coll}, it retains an arbitrary $u$-dependence.
The same computation in arbitrary dimension shows that the contribution from this region scales like $r^{-\frac{D-2}{2}}$.

Altogether, for the leading order from both regions one has
\begin{equation}\label{leading cartesian}
    H_{\m \n}  \sim \frac{1}{r}\ve_{\m \n} (\vz\,) \left(\eta^{\text{reg}}(\vec{z}\,)+\eta^{ \text{coll}} (u,\vz\,)\right),
\end{equation} 
with $ \eta^{\text{reg}}(\vz\,), \,   \eta^{\text{coll}}  (u,\vz\,)$ scalar functions encoding contributions from $H_{\mu \nu}^{\text{reg}}$ and $H_{\mu \nu}^{\text{coll}}$, respectively.
As a consequence, in force of \eqref{Bondi parameter} and \eqref{leading cartesian} one finds that
\begin{equation} \label{Hradial}
    H_{r \n} = \frac{\pr x^\m}{\pr r} H_{\m\n} =\frac{q^\m(\vec{z}\,)}{q^0(\vz\,) } H_{\m \n} 
\end{equation}
implying that the radial components of the DC field fall off faster than $1/r$, owing to the transversality of $\ve_{\m \n} (\vz\,)$ in \eqref{leading cartesian}. Note that, due to the transverse projection, our discussion of spin-1 and spin-2 fields does not include the information on the Coulombic fields that determine the total global charges. These considerations will be important for \eqref{falloff_h} and \eqref{falloff_B} below.

\section{Asymptotic single-copy parameters} \label{dc_parameters_null}

Let us now move to the analysis of the behaviour of the gauge parameters at null infinity. We will start by determining the conditions on the large-$r$ expansion of the single-copy parameters which allow one to describe spin-one asymptotic symmetries. In the next section, we will then proceed to identify the implications on the corresponding double-copy gauge parameters defined in \eqref{DC parameters}.

 In Lorenz gauge one has 
\begin{equation} \label{wave_epsilon}
    \Box \, \e^a = 0 = \Box \, \tilde{\e}^{a'}
\end{equation}
so that the general form of the spin-one parameters is 
\begin{equation} \label{vector_parameters}
	\begin{aligned}
  {\e} (x) = \int_k e^{ik\cdot x} {a}^{(\e)}(k)\,, \qquad
  \tilde{\e} (x) = \int_k e^{ik\cdot x} \tilde{a}^{(\e)}(k)\,.
	\end{aligned}
\end{equation}

The very existence of spin-one asymptotic symmetries in this gauge requires the parameters to admit a polyhomogeneous expansion of the form\footnote{Here and in what follows we denote with $f^{(n)}$ the component of ${\cal O} (r^{-n})$ in the radial expansion of a given function $f$, and use a  different letter for coefficients multiplying logarithmic terms, if any.}
\begin{equation}\label{poly e}
\e (r, u, \vz\,) = \sum_{n=0}^{\infty} \dfrac{\e^{(n)}  (u, \vz\,)}{r^n} + \sum_{n=1}^{\infty}\l^{(n)} (u, \vz\,) \dfrac{\log r}{r^n}\, .
\end{equation}
Indeed, if the logarithmic series in \eqref{poly e} were not included, the wave equations \eqref{wave_epsilon} would imply the conditions
\begin{equation}
    \cD^2 \, \e^{(0)} = 0 = \cD^2\,  \tilde{\epsilon}^{(0)}\, , 
\end{equation}
that would force the leading order parameters to be constant and would thus set to zero the putative asymptotic-symmetry parameters. The same mechanism holds for vector parameters as well, and it will be instrumental to grant the existence of asymptotic symmetries for the graviton and the two-form components of the DC field \cite{Campiglia:2016hvg,Campoleoni:2019ptc}.

Thus, the scaling behaviour of the Fourier coefficients in \eqref{vector_parameters} must be such to grant not only the existence of the ${\cal O} (1)$ coefficients, that can only come from the regular region in the large-$r$ analysis of \eqref{vector_parameters}, but also the first subleading terms to scale like $\frac{\log r}{r}$, while in addition being linear\footnote{The logarithmic tail in \eqref{vector_parameters} will generate  logarithmic falloffs on  some field components as well, although in a pure-gauge sector that won't affect any gauge-invariant observable. Even if included directly in the field expansion, however, it is possible to show that they would not change {\it e.g.} the finiteness of the energy.} in $u$. With reference to the parameterisation \eqref{Bondi parameter}, an  expansion compatible with \eqref{poly e} is
\begin{equation}\label{expansion O(1)}
\begin{split}
 a^{(\e)} (\o q(\vw))\sim &\frac{i}{\pi^2} \o^{\b-2}\big( 2\pi b^{(\e)}_0 (\vw) 
 - \log \o\, a^{(\e)}_0 (\vw)\big),
 \end{split}
\end{equation}
where all functions are real\footnote{The most general leading-order expansion would include an imaginary part for $b^{(\e)}_0$ which, in its turn, would generate an additional contribution to the asymptotic symmetries.}. Let us discuss the two contributions separately. 

From $ b^{(\e)}_0 $ we find to leading order
 \begin{equation}
\e_{\text{\textit{glob}}} = 2\frac{\G(\b)}{\pi \, r^{\b}}  \sin\left(\frac{\pi}{2} \b\right) \int d^2\vw \,b^{(\e)}_0 (\vw)\, |\vw-\vz\,|^{-2\b} ,
\end{equation}
which, in the limit $ \b\rightarrow 0 $, implies
\begin{equation}\label{global eps}
	\e_{\text{\textit{glob}}} =  \int d^2\vw \,b^{(\e)}_0 (\vw) ,
\end{equation}
which is well defined for integrable $ b^{(\e)}_0 (\vw) $ and does not depend on $\vz$, consistently with the fact that without 
admitting a polyhomogeneous expansion in \eqref{poly e} one can only  find a {\it constant} leading order parameter. 

From $a^{(\e)}_0$, we find instead
\begin{equation}\label{e_as}
		\begin{split}
		    \e_{\text{\textit{as}}} =&  \frac{1}{2\pi} \left(\log r + \g_E \right)\int d^2\vw \; a^{(\e)}_0 + \frac{1}{2\pi} \int d^2\vw\; a^{(\e)}_0\, \log\left(\frac{|\vw-\vz\,|^{2}}{1+|\vz\,|^{2}}\right) + \cO(\b).
		\end{split}
\end{equation}
where $\g_E$ is the Euler--Mascheroni constant.
In particular, consistency with \eqref{poly e} requires the  $\log r$ term not to be present, a condition that holds   whenever $a^{(\e)}_0$ is a total derivative. 
We will consider
\begin{equation}\label{ansatz_double_lap}
    a^{(\e)}_0 (\vw)= \D^2 f (\vw),
\end{equation}
where $\D$ is the $\mathbb{R}^2$ Laplacian. 
Due to \eqref{ansatz_double_lap} the first integral in \eqref{e_as} cancels while, upon  integrating by parts  the second one, one finds, for $\b\to 0$
\begin{equation}\label{epsilon0}
	\e_{\text{\textit{as}}} = \D f (\vz\,),
\end{equation}
thus showing how one can obtain, together with global transformations, asymptotic gauge parameters retaining  a nontrivial dependence on the angular coordinates. Let us observe that the first subleading term resulting from the ansatz \eqref{expansion O(1)}, is indeed of order $\frac{\log r }{r}$ with coefficient
\begin{equation}
\l^{(1)} = \frac{1}{2}\,u\, \cD^2 \e_{\text{\textit{as}}}
\end{equation}
 linear in $u$, consistently with ref. \cite{Campoleoni:2019ptc}. 

\section{Double copy of asymptotic symmetries} \label{dc_asymptotica}

Let us now analyse the asymptotic behaviour of the DC parameters $\a_\m$ and $\tilde{\a}_\m$ given in \eqref{DC parameters}, as determined by the behaviour their single-copy constituents discussed in the previous section. In force of the single-copy Lorenz gauge they satisfy 
\begin{equation} \label{conditions alpha}
	\Box \, \a_\m = 0,\, \nabla^\m \a_\m = 0 \qquad \mbox{and} \qquad \Box \, \tilde{\a}_\m = 0, \, \nabla^\m \tilde{\a}_\m = 0\, ,
\end{equation}
so that the general form of their Minkowskian components is 
\begin{equation} \label{DC fourier}
	\begin{aligned}
	  \a_\m(x) \!=\!\! \int_k e^{ik\cdot x} \frac{a^{(\e)}\tilde{a}^{(A)} }{a^{(\F)}}\,\tilde{\ve}_\m\,,\qquad 
	  \tilde{\a}_\m(x) \!=\!\! \int_k e^{ik\cdot x} \frac{{a}^{(A)}\tilde{a}^{(\e)} }{a^{(\F)}}\,\ve_\m\,.
	\end{aligned}
\end{equation}
We thus see that the $\o$-scaling of the Fourier coefficients,
\begin{equation}
a^{(\a)} = \frac{a^{(\e)}\tilde{a}^{(A)} }{ a^{(\F)}}\,, \qquad \tilde{a}^{(\a)} = \frac{a^{(A)}\tilde{a}^{(\e)}}{a^{(\F)}}\,,
\end{equation}
is the same scaling as that of $a^{(\e)}$ and $\tilde{a}^{(\e)}$, since in our setup $a^{(A)}$, $\tilde{a}^{(A)}$ and  $a^{(\F)}$ all scale like $\o^{-1}$, as specified in \eqref{single_copy_scaling}. For instance for $a^{(\a)}$ we have
\begin{equation}
	a^{(\a)} (\omega q({\vw})) \sim \frac{i}{\pi^2}\o^{\b-2} \big(2\pi b^{(\a)}_0(q(\vw) - \log \o\,  a^{(\a)}_0(q(\vw)) \big),
\end{equation}
with
\begin{equation} \label{a^a}
    b^{(\a)}_0 = \frac{\tilde a_{0}^{(A)}}{a_{0}^{(\F)}} b_0^{(\e)},  \qquad a^{(\a)}_0 = \frac{\tilde a_{0}^{(A)}}{a_{0}^{(\F)}} \D^2 f  .
\end{equation}
where we took our choice \eqref{ansatz_double_lap} for $a^{(\e)}_0$ into account. On closer inspection, however, $a^{(\a)}_0$ in \eqref{a^a} is not a total derivative. In this sense, evaluating the leading order in \eqref{DC fourier} by means of the counterpart of  \eqref{e_as}, one would seemingly have to retain also the contributions corresponding to the first two terms to the r.h.s. of \eqref{e_as} and thus, in particular, a term diverging as  $\cO(\log r)$.  

However, let us recall that, for asymptotic fields generated by sources moving in the bulk, the form of the leading contribution for soft momenta $q$ is fixed by soft theorems to be the sum of terms of the following type \cite{Strominger:2014pwa,Saha:2019tub,Sahoo:2020ryf,Sahoo:2021ctw}
\begin{equation}\label{soft terms}
    a_0^{(A)} = e\, \frac{\varepsilon\cdot p}{p\cdot q}\,,\qquad
    a_0^{(\tilde A)}=\ \tilde {e}\, \frac{\tilde{\ve}\cdot p}{p\cdot q}\,, \qquad
    a_0^{(\Phi)} =\frac{g^\Phi}{p\cdot q}\,,
\end{equation}
each associated to the motion of a given background particle of hard momentum $p$, with $e$, $\tilde{e}$ and $g^\Phi$ the corresponding couplings and  where we choose $\varepsilon_i^\mu (\vz\,)=(z_i,\delta_{ij},-z_i)=\tilde \varepsilon_i^\mu(\vz\,)$, which is equivalent to set the reference vector $y$ introduced in \eqref{momentum_pol} to $y^\m = (-1,0,0,1)$.  From the DC perspective \cite{Bjerrum-Bohr:2010pnr, DiVecchia:2017gfi}, where at each vertex the particles in the gravitational multiplets are interpreted as products of single-copy constituents, upon further choosing $g^\Phi =  \frac{e\tilde e}{\sqrt{8\pi G}}$, with $G$ denoting the gravitational constant, one finds 
\begin{equation} \label{grav_coupling}
    a_0^{(H)} = \sqrt{8\pi G}\ \frac{\ve\cdot p \; \tilde{\ve} \cdot p}{p\cdot q}.
\end{equation}
This observation has an immediate implication on the structure of the ratios in \eqref{a^a}, that have to take the form
\begin{equation}\label{soft dc}
	\frac{\tilde a_0^{(A)}}{a_0^{(\F)}} =  \frac{\sqrt{8\pi G}}{e} \; \tilde\ve \cdot p\, .
\end{equation}
Making use of \eqref{soft dc} and \eqref{ansatz_double_lap}  one finds that the dangerous term eventually cancels, since 
\begin{equation}
    \frac{\log r + \g_E}{2\pi}\int d^d\vw \, (\D^2 f) \; \tilde\ve \cdot p \; \tilde\ve_\m = 0 
\end{equation}
because $\D^2(\tilde\ve \cdot p \; \tilde \ve_\m )=0 $,  which motivates our choice  \eqref{ansatz_double_lap}.

Let us highlight two main features of the so-derived asymptotic behaviour of the DC parameters: 

\begin{description}
\item[(i)] They keep  a nontrivial angular dependence at $ \cO(1) $ for large $r$. In terms of single-copy inputs we find, performing the integral in \eqref{DC fourier},
\begin{equation}\label{DC par expression}
    \a^{(0)\m} (\vz\,) = \a^\m_{\text{\textit{glob}}} + \D f \, \tilde\Psi^\m  + f \D \tilde\Psi^\m - 2 \pr_i f \pr^i \tilde\Psi^\m,
\end{equation}
where we defined 
\begin{equation}
    \tilde \Psi^\m = \frac{\tilde a_0^{(A)}}{a_0^{(\F)}} \tilde \ve^\m = \frac{\sqrt{8\pi G}}{e}  \; \tilde\ve\cdot p \; \tilde\ve^\m . 
\end{equation}

\item[(ii)]
They inherit from their single-copy constituents a polyhomogeneous expansion of the form  
\begin{equation} \label{polyalpha}
\a^{\m} = \a^{(0){\m}} (\vz\,) + {\cal O}\left(\frac{\log r}{r}\right) \, .
\end{equation}
\end{description}
The Bondi components of $\a^{\m}$ are related to the Minkowski ones, here denoted as $\a^{\m}_M$, by 
\begin{equation}\label{bondi leading par}
		\a^u = - \hat q \cdot \a_M, \qquad
	\a^r  = \hat{n}^I \a_M^I, \qquad
		\a^i  = \frac{1}{r} (\cD^i \hat{n}^I )\a^I_M\, ,
\end{equation}
with $I = 1, 2, 3$ and 
\begin{equation}
\hat{q}^\m (\vz\,)= \frac{q^\m(\vz\,)}{q^0(\vz\,)} = (1,\hat{n}^I(\vz\,)),
\end{equation} 
while $\cD_i$ denotes the covariant derivative on the sphere. They satisfy 
\begin{equation} \label{alpha constraint}
\cD^2\,  \a^{(0)u}+2  \a^{(0)r} +2 \cD_i \a^{(1)i}= - \hat q \cdot \cD^2\,\a_M^{(0)} = 0 \,  
\end{equation}
as a consequence of \eqref{conditions alpha}, where $\cD^2 = \cD_i\cD^i$  and we used $ \cD^2 \hat n^I = -2 \hat n^I $. For the ensuing discussion it is useful to  parameterise the solutions of \eqref{alpha constraint} in terms of a scalar function $\Theta(\vz\,) $ and vector on the sphere $\Sigma^i(\vz\,)$  as follows:
\begin{equation} \label{alpha parametrization}
\begin{split}
\a^{(0)u} (\vz\,)  &= \Theta (\vz\,), \\\a^{(1)i}(\vz\,) &= - \cD^i \Theta (\vz\,)  + \Sigma^i (\vz\,) ,  \\
\a^{(0)r} (\vz\,) &=  \frac{1}{2} \cD^2\, \Theta (\vz\,)  - \cD_i \Sigma^i (\vz\,) \, .
\end{split}
\end{equation}
At this point, we collected all the results needed in order to analyse the asymptotic symmetries of the  multiplet subsumed in $H_{\m\n}$ From the Lagrangian \eqref{lagrangian} one can see that $H^S_{\m\n}$ and $H^A_{\m\n}$ decouple, which makes it sensible to analyse their asymptotic properties separately. In particular, our analysis for the two-form sector will entail some novel considerations with respect to those available in the literature, and for this reason we shall discuss it in more detail. 

Given that  the asymptotic charges involve appropriate components of the curvatures, let us recall the definition of the DC curvature given in \cite{Anastasiou:2014qba}:
\begin{equation}
\begin{split}
R_{\m\n\r\s} &= -\frac{1}{2} F_{\m\n} \star \tilde{F}_{\r\s}  = R^S_{\m\n\r\s} + R^A_{\m\n\r\s}\, ,
\end{split}
\end{equation}
involving  the linearised Riemann tensor for $H^S_{\m\n}$ together with a gauge invariant  combination of derivatives of  $H^A_{\m\n}$, whose Minkowskian components both look:
\begin{equation}
R^I_{\mu\nu\rho\sigma} = \frac{1}{2} (\partial_{\mu} \partial_{\sigma} H^I_{\nu\rho} - \partial_{\nu} \partial_{\sigma} H^I_{\mu\rho} + \partial_{\nu} \partial_{\rho} H^I_{\mu\sigma} -\partial_{\mu} \partial_{\rho} H^I_{\nu\sigma})\, ,
\end{equation}
with $I = A, S$. More explicitly, it is useful to recognise that 
\begin{equation}\label{RA}
R^A_{\mu\nu\rho\sigma}= \frac{1}{2} \left( \partial_{\mu}\mathcal{H}_{\nu\rho\sigma}\, - \, \partial_{\nu}\mathcal{H}_{\mu\rho\sigma} ,  \right)
\end{equation} 
with 
\begin{equation}
{\mathcal{H}}_{\a\b\g} = \partial_{\a} H^A_{\b\g} + \partial_{\g} H^A_{\a\b} + \partial_{\b} H^A_{\g\a}\, 
\end{equation}
the field strength for the two-form $H^A_{\m\n}$.
\subsection{Supertranslations and scalar charges} \label{dc_parameters_null_symm}

The gauge parameter of $H^S_{\m\n}$ is 
\begin{equation} \label{graviton parameters}
	\xi^\m = \frac{1}{2}(\a^\m + \tilde{\a}^\m)\, 
\end{equation}
and satisfies 
\begin{equation}
\Box\xi^\m = 0\,,\qquad \nabla \cdot \xi = 0\,,
\end{equation}
due to \eqref{conditions alpha}. The asymptotic surface charges  correspond to the fixed$-u$, large$-r$
limit of 
\begin{equation}\label{QH}
	Q^S = - r^3\int dzd\bz \g_{z\bz}\left(\xi^u R^S_{u r u r} + \xi^iR^S_{irur}\right),
\end{equation}
where, due to \eqref{DC multiplet}, one can identify two contributions to the curvature 
\begin{equation}
	R^S_{\m\n\r\s} = R^h_{\m\n\r\s}+R^\vf_{\m\n\r\s},
\end{equation}
associated to the  graviton $h_{\m\n}$ and to the scalar $\vf$, respectively.
 In particular, because of \eqref{LorenzH}, the graviton $h_{\m\n}$ is in de Donder gauge, while the condition $\nabla \cdot \xi = 0$ is consistent with the  scalar degree of freedom being encoded in the equation for the trace of $H^S_{\m\n}$ \cite{Ferrero:2020vww, Francia:2010qp, Campoleoni:2012th}.
 Supertranslations in de Donder gauge were discussed in \cite{Avery:2015gxa, Campiglia:2016efb, Himwich:2019qmj} and in the remainder of this section we will show how we recover the corresponding results from the DC.
 
 The falloffs on $H_{\m \n}$ imply the following scalings:
\begin{equation}\label{falloff_h}
\begin{split}
	\vf, h_{uu}, h_{ur}, h_{ri} & =\cO\left( \tfrac{1}{r} \right),\qquad	h_{rr} =\cO\left( \tfrac{1}{r^2} \right), \\
	h_{ui}, h_{z\bz} & =\cO\left(1\right),\qquad 	h_{zz},h_{\bz\bz} =\cO\left( r \right) ,
\end{split}
\end{equation}
while the leading Bondi components of $\xi^{\m}$ satisfy \eqref{alpha constraint} and \eqref{alpha parametrization},
with the substitutions
\begin{equation} \label{Theta_Sigma_symm}
    \begin{split}
\Theta (\vz\,) \quad \rightarrow     \quad 
T(\vz\,)\; &= \frac{1}{2} \left(\Theta (\vz\,) + \tilde \Theta(\vz\,)\right) , \\     
\Sigma (\vz\,)  \quad \rightarrow  \quad 
S^i(\vz\,) &= \frac{1}{2} \left(\Sigma^i (\vz\,) + \tilde \Sigma^i (\vz\,)\right) .
\end{split}
\end{equation}
It is possible to perform small gauge transformations employing $S^i$ setting to zero  $h^{(1)}_{ur}$ and $h^{(2)}_{rr}$.
The residual parameters are then further constrained by 
$
	\hat q \cdot \cD_i \xi = 0\, ,
$
which suffices to fix the correct cross-dependences among the components of $\xi^{\m}$, namely
\begin{equation} \xi^{(1)i} = - \cD^i T(\vz\,), \quad \quad   	\xi^{(0)r} = \frac{1}{2} \cD^2 \, T(\vz\,)\, .
\end{equation}
This identifies the leading-order asymptotic symmetries of the DC graviton as BMS supertranslations. The corresponding charge receives contribution only from the first term in \eqref{QH}, on account of \eqref{falloff_h}, and has the form
\begin{equation} \label{supertranslation charge}
Q^h  = 
- \int dzd\bz \g_{z\bz} T (z,\bz) R^{h (3)}_{u r u r},
\end{equation}
with 
\begin{equation}
R^{h (3)}_{urur} = -\frac{1}{2}\cD^i\cD^jh^{(-1)}_{ij}\, .
\end{equation}

In addition, from both terms in \eqref{QH} one finds that the DC builds infinitely-many asymptotic charges for the scalar field of the form
\begin{equation}
		Q^{\vf} =-\frac{1}{2} \int dz d\bz \g_{z\bz} (\cD^2 + 1) T(z,\bz) \vf^{(1)}\, ,  
\end{equation}
 which reproduce the asymptotic  charges first proposed for scalar fields in \cite{Campiglia:2017dpg}, with the identification of the corresponding smearing function with $(\cD^2 + 1) T(z,\bz)$. 

\subsection{Two-form asymptotic symmetries} \label{dc_parameters_null_asymm}

In the present section we denote the two form as 
\begin{equation}
 H^A_{\m\n} = B_{\m\n}\, .
\end{equation}
The parameter of $B_{\m\n}$ in \eqref{DC multiplet} is defined as 
\begin{equation} \label{two form parameters}
\L^\m = \frac{1}{2}(\a^\m - \tal^\m) 
\end{equation}
and satisfies 
\begin{equation}
\Box\L^\m = 0,\qquad  \nabla \cdot \L = 0,
\end{equation}
where the latter condition is tantamount to a choice of gauge-for-gauge, implemented via \eqref{gfg}. The asymptotic charge corresponding to \eqref{QH} emerges from the $r \rightarrow + \infty$ limit of 
\begin{equation} 
	Q^{B}  =  - r^3 \int dz d\bz \g_{z\bz} \! \left(\L^u  R^A_{u r u r}\! + \! \L^i  R^A_{i r  ur}\right)\, ,
\end{equation}
where in this case the first contribution vanishes identically. Thus, its very  existence depends on the presence of a term of ${\cal O} (\frac{1}{r^3})$ in $\L^i  R^A_{i r  ur}$. The latter, in its turn, originates from
$ \L^{i (1)}  R^{A (2)}_{i r  ur}$
with 
\begin{equation}
  R^{A (2)}_{i r  ur} = {\cal H}^{(1)}_{iur} = \partial_u B^{(1)}_{ri} = - {\cal D}^j B^{(-1)}_{ij} \, ,
\end{equation}
where in the last equality we used the equations of motion. Altogether, the tentative asymptotic charge is
\begin{equation}\label{QB}
	Q^{B}  =   \int dz d\bz \g_{z\bz}\,  \L^{i (1)}\ {\cal D}^j B^{(-1)}_{ij} \, ,
\end{equation}
and we need to determine whether in our DC setup there is room for a nonvanishing coefficient $\L^{i (1)}$.
From \eqref{polyalpha} we obtain
\begin{equation} \label{polyL}
    \begin{split}
		&\L^u = \sum_{n=0}^{\infty}\frac{\L^{(n)u}}{r^n}+\sum_{n=1}^{\infty}\l^{(n)u}\frac{\log r}{r^n},\\
		&\L^r = \sum_{n=0}^{\infty}\frac{\L^{(n)r}}{r^n}+\sum_{n=1}^{\infty}\l^{(n)r}\frac{\log r}{r^n},\\
		&\L^i = \sum_{n=1}^{\infty}\frac{\L^{(n)i}}{r^n}+\sum_{n=2}^{\infty}\l^{(n)i}\frac{\log r}{r^n},
	\end{split}
\end{equation}
where in particular the leading coefficients are related by 
\begin{equation}\label{log corr}
	\begin{split}
 & \cD^2 \L^{(0)u} + 2{\L^{(0)r}}+2\cD \cdot {\L^{(1)}}=0\, , \\
		&2\pr_u{{\l^{(1)r}}}=(\cD^2-2){{\L^{(0)r}}}-2 \cD\cdot {{\L^{(1)}}}\, ,\\
		&2\pr_u{{\l^{(2)i}}}=(\cD^2-1){{\L^{(1)i}}}+2\cD^i {{\L^{(0)r}}}\, ,
	\end{split}
\end{equation}
as a consequence of the Lorenz gauge condition. We parameterise the corresponding solutions as follows
\begin{equation} \label{leading_Lambda}
\begin{split}
	{\L^{(0)u}} & = \t(\vz\,)\, ,\\ {\L^{(1)i}} & =- \cD^i\t(\vz\,)+ \s^i(\vz\,)\, ,\\  {\L^{(0)r}}     &= \frac{1}{2}\cD^2 \t(\vz\,)-\cD\cdot\s(\vz\,)\, .
\end{split}
\end{equation}
having defined
\begin{equation} \label{Theta_Sigma_asymm}
    \begin{split}
    \t(\vz\,) & = \frac{1}{2} \left(\Theta (\vz\,) - \tilde \Theta(\vz\,)\right), \\     
 \s^i(\vz\,) & = \frac{1}{2} \left(\Sigma^i (\vz\,) - \tilde \Sigma^i (\vz\,)\right).
\end{split}
\end{equation}
with $\t(\vz\,)$ and $\s^i (\vz\,)$ providing the counterparts of \eqref{Theta_Sigma_symm}. From the second of \eqref{leading_Lambda} we see that in principle there are two types of contributions to $Q^B$, depending on  $\t(\vz\,)$ and on $\s^i (\vz\,)$, respectively. The contribution to the charge involving 
$\t (\vz\,)$, however, actually vanishes:
\begin{equation} \label{QBtau}
\begin{split}
    Q_{\t}^B =  \int dz d\bz \, \g^{z\bar{z}}(\cD_{\bar{z}}\cD_z \t -  \cD_z \cD_{\bar{z}}\t)B_{z\bar{z}}^{(-1)} = 0\, .
\end{split}
\end{equation}
This outcome can be understood upon recognising that variations involving $\L^{(1)i} = -\cD^i \t $ represent gauge-for-gauge transformations that do not affect the free data encoded in $B^{(-1)}_{ij}$.
Thus, it is the presence of a non-trivial $\s^i (\vz\,)$ that allows one to identify the asymptotic charge of the two-form:
\begin{equation} \label{two-form charge}
{Q}^B_{\s} = \int dzd\bz \, \g^{z\bar{z}}(\cD_{\bz}\s_{z}-\cD_{z} \s_{\bz}) B_{z\bar{z}}^{(-1)}  \, .
\end{equation}	
Let us observe that the parameters $\t(\vz\,)$ and $\s^i (\vz\,)$ play a role which is somehow reversed w.r.t. their gravitational counterparts $T(\vz\,)$ and $S^i (\vz\,)$ defined in \eqref{Theta_Sigma_symm}, with the 
supertranslation parameter $T(\vec{z})$ providing the relevant asymptotic symmetry and $S^i (\vz\,)$ encoding small gauge transformations.

A few comments are in order:
\begin{itemize}
 \item The field coefficient $B_{z\bar{z}}^{(-1)}$ appearing in the charge \eqref{QB} is consistent with the single-copy induced falloffs of the two-form components,
\begin{equation}\label{falloff_B}
    \begin{aligned}
			&B_{ur} = \cO\left(\tfrac{1}{r^2}\right)\,,\qquad
			B_{ui} = \cO\left(1\right)\,,\\
			&B_{ri} = \cO\left(\tfrac{1}{r}\right)\,,\qquad\,\,\,
			B_{ij} =\cO\left(r\right)\, ,
   \end{aligned}
\end{equation}
where we also took into account that the Lorenz gauge together with the equations of motion set to zero $B_{ur}^{(1)}$ and $B_{ri}^{(0)}$. (See \eqref{Hradial}.)
 \item The polyhomogeneous expansion \eqref{polyL}, however, induces partial violations of the falloffs \eqref{falloff_B}. Indeed, if $\s^i \neq 0$ then 
\begin{equation} \label{polyB}
\begin{split}
\d B_{ur} = \cO \left(\tfrac{1}{r}\right),\quad\; \d  B_{ui} = \cO(\log r),\quad\;
\d B_{ri} =\cO(1),\quad\;  \d  B_{ij} = \cO(r)\,,    
\end{split}
\end{equation}
and therefore the falloffs are generically not preserved except for $B_{ij}$, which encodes the free data. The same mechanism is at work for the spin-one and spin-two cases in Lorenz and de Donder gauges, respectively, where polyhomogeneous scalings of the parameters were found to be necessary in order for asymptotic charges to exist. One possible attitude is to intepret the violations of \eqref{falloff_B} as physically innocuous, since they affect only pure-gauge components  of the field and as such do not alter any physical observables \cite{Campoleoni:2019ptc}.  Alternatively, one may also envisage the possibility to generalise the falloffs of the fields from the very start, as suggested in \cite{Himwich:2019qmj, Peraza:2023ivy}, possibly in order to take the presence of matter into account. As an indication supporting this option one can show that even with falloffs modified as in \eqref{polyB} the energy flux through null infinity would be finite anyway.
 \item The asymptotic symmetries at null infinity of two-forms in $D=4$ Minkowski space were analysed in Lorenz gauge in \cite{Campiglia:2018see} and in radial gauge in \cite{Francia:2018jtb}, to the purpose of identifying dual counterparts of the conserved asymptotic charges for scalars found in \cite{Campiglia:2017dpg}.
See also \cite{Afshar:2018apx, Henneaux:2018mgn} for analyses at spatial infinity. In both \cite{Campiglia:2018see} and \cite{Francia:2018jtb}, however, the scalings of the parameters were chosen so as to always {\it preserve} the (putative) falloffs of the fields. In the radial gauge chosen in \cite{Francia:2018jtb}, the observed outcome was the absence of an order-zero asymptotic charge. In the Lorenz-gauge analysis of \cite{Campiglia:2018see}, which is closer to our present perspective, the parameter falloffs in particular had to preserve the condition
\begin{equation} \label{B_ui}
B_{ui} = \cO(1)\, ,
\end{equation}
which, as shown above, is incompatible with the polyhomogeneous scalings \eqref{polyL}. If one were to remove the logarithmic terms  in \eqref{polyL}, however, the actual falloffs for the fields would eventually match up to those in \eqref{falloff_B}, as one can recognise by taking the equations of motion into account, while the corresponding equations on the leading components of the parameter would imply that only the $\t$ contribution to $\L^{i (1)}$ in \eqref{leading_Lambda} would be different from zero. For the latter however, as shown in \eqref{QBtau}, the tentative charge actually vanishes. To summarise, according to our findings the only way to preserve the falloff \eqref{B_ui} is by restricting  the gauge parameters in such a way that the  charge itself vanishes, and therefore does not identify an asymptotic symmetry.
\end{itemize}
The charges  \eqref{two-form charge} should provide the dual counterparts of the asymptotic charges for scalar fields found in \cite{Campiglia:2017dpg}.

\section{Conclusions} \label{conclusions}
In this work, we presented a concrete and general incarnation of the DC of asymptotic symmetries, whereby electromagnetic large gauge transformations ``square'' to BMS supertranslations for the graviton and to asymptotic symmetries for the Kalb--Ramond two-form, while also providing an explanation to the existence of infinitely-many asymptotic charges for the scalar field. Our approach consisted in performing the asymptotic expansions of the on-shell fields by means of the method of regions, which allowed us to deduce the falloffs of the double-copy fields from those of their single copy constituents under suitable assumptions. A similar mechanism works for the corresponding asymptotic symmetry parameters. The appearance of terms involving $\log r$ at subleading orders, which is a common feature of asymptotic expansions performed in Lorenz and de Donder gauges, motivated us to revisit the study of the asymptotic symmetries and charges for the double-copy fields, in particular for the two-form. We found that the logarithms ``predicted'' by the convolutional double copy in this way are precisely those needed to allow for the existence of nontrivial asymptotic symmetries and charges.

There are several possible generalisations of our work that could be considered in the future. As observed, the study of the DC for asymptotic symmetries in the convolutional approach is tightly linked to soft theorems and it would be interesting to better investigate this connection. Let us mention that an alternative route to asymptotic symmetries, and thus to their double copy, is also offered by the representation of asymptotic symmetries in terms of OPE expansions of celestial conformal field theory amplitudes \cite{Banerjee:2022hpo, Banerjee:2022lnz}. Moreover, while in this work we have only considered the leading terms in the expansion of fields and parameters, one could consider higher-order terms as well as non-linear corrections, exploring their possible connection with subleading soft theorems. In particular, while here we have focused on reproducing BMS supertranslations from the large gauge transformations of spin one particles, it is natural to wonder whether gravitational superrotations \cite{Barnich:2009se,Campiglia:2016efb} can also be encompassed by this framework.

Further potential developments include the generalisation of our approach to different, yet related, contexts. The study of asymptotic symmetries and the associated soft theorems has been generalised to (Anti) de Sitter backgrounds \cite{Esmaeili:2019mbw,Fiorucci:2020xto,Esmaeili:2021szb,Campoleoni:2023eqp}, for which the convolutional DC dictionary has recently been investigated in \cite{Liang:2023zxo}. 
Another enticing perspective is to investigate
asymptotic symmetries of higher spin gauge fields \cite{Campoleoni:2017mbt,Campoleoni:2020ejn} formulated as convolutional double copies of lower-spin building blocks\footnote{See \cite{Ponomarev:2017nrr} for a chiral higher-spin DC and \cite{Didenko:2022qxq} for a Kerr-Schild-type one.}. Moreover, since the convolutional DC can be equivalently formulated in any number of spacetime dimensions, it is certainly possible to imagine that our work and the generalisations thereof could be also explored beyond $D=4$. 

\appendix
\section{Coordinate conventions}\label{coordinates}
We adopt the mostly plus convention for the metric. Bondi coordinates in $D=4$ are defined as $(u, r, z^i)$, with $r$ denoting the radial coordinate, $u$ the retarded time and $z^i\; (i=1,2)$ the angular coordinates, which we parameterise either as stereographic coordinates  
\begin{equation}
 z = e^{i\f} \cot\frac{\th}{2} \qquad \mbox{and} \qquad  \bar{z} = e^{-i\f} \cot\frac{\th}{2}\, ,   
\end{equation}
with $\theta \in [0, \pi]$ and $\phi \in [0, 2 \pi)$, or in terms of a real vector 
\begin{equation}
\vz  = (z_1 = \text{Re} z, z_2 = \text{Im} z) \, .
\end{equation} The Minkowski metric reads
\begin{equation}
    ds^2=-du^2-2du\, dr + r^2 \gamma_{ij}dz^i\,dz^j,
\end{equation} 
where $\gamma_{ij} $ is the unit metric on the two-sphere, which is anti-diagonal in terms of $(z,\bz)$, with $\gamma_{z\bar{z}} = \frac{2}{(1+z\bar{z})^2}$, while it is diagonal when using $\vz$, namely $\g_{ij} = \frac{2}{(1+|\vz|^2)^2}\d_{ij} $.
For the Minkowski coordinates $x^\m$ we also adopt the following parameterization:
\begin{equation} \label{Bondi parameter_app}
\begin{split}
	x^\m = u t^\m + \frac{2r}{1+|\vz\,|^2}q^\m(\vz\,)\, ,
\end{split}
\end{equation}
where $t^\m=(1,0,0,0)$ and  
\begin{equation}
q^\m(\vz\,) = \frac{1}{2}\left(1+|\vec{z}\,|^2,2z_1,2z_2, 1-|\vec{z}\,|^2\right).
\end{equation}
We denote with $\cD_i$ the covariant derivative with respect to $\g_{ij}$ and with $\cD^2$ the corresponding Laplace operator: $\cD_i \cD^i= \cD^2$, where case by case we take care of clarifying the type of $z-$coordinates that we use in the various situations. At the same time, we use $\pr_i$ to denote the partial derivative with respect to $z^i$ and we define $\D:=\pr_i\pr^i$.

For the momenta we use Bondi-like coordinates, namely $(\m,\o, w^i)$, where $\m = k^0-|\vk|$ is an analog of the retarded time $u$ and $w^i$ are the angular components of the momentum, which can be parameterised again either as stereographic coordinates $(w,\bar w)$ or in terms of a real vector $\vw$. However, $\o$ is not the standard frequency but 
a convenient re-definition thereof useful to employ in various integrals, as it will become clear soon. In particular, we use $\o = \tfrac{|\vk|}{q^0(\vw)}$ so that the counterpart of \eqref{Bondi parameter_app} is:
\begin{equation}
	k^\m  = \mu t^\m + \o q^\m(\vw)\,,
\end{equation}
In the $(\mu, \o, \vec{w})-$coordinates the metric reads
\begin{equation}\label{k_metric}
    dk^2 = -d\m^2 -(1+|\vw|^2)\,d\m d\o -2\o  w_i\, d\m dw^i +\o^2 \d_{ij}dw^idw^j
\end{equation}
and therefore the unit metric on the two sphere is now simply $\g_{ij}(\vw) = \d_{ij}$. Hence, the covariant derivative w.r.t.\ $\g_{ij}(\vw)$ is an ordinary partial derivative $\pr_i$, providing great simplifications when performing integration by parts.
Polarisation vectors in momentum space can be defined as \cite{Donnay:2022ijr}
\begin{equation} \label{momentum_pol}
    \ve^\m_i = \pr_i \left( \frac{q^\m}{y\cdot q} \right),
\end{equation}
with $y$ a null reference vector.
\paragraph*{\bf{Acknowledgments.}} D.F. is grateful to F. Manzoni for discussions. C.~H. is supported by UK Research and Innovation (UKRI) under the UK government’s Horizon Europe funding guarantee EP/X037312/1
\bibliographystyle{utphys} 

\end{document}